\documentclass[aps,twocolumn,superscriptaddress,floatfix,secnumarabic,balancelastpage,amsmath,amssymb,nofootinbib,
superscriptaddress,prb,10pt]{revtex4-2}
\usepackage[utf8]{inputenc}
\usepackage{soul}
\usepackage{tabularx}
\usepackage{siunitx}
\usepackage{array}
\usepackage{braket}
\usepackage{graphicx}
\usepackage{longtable}
\usepackage{textcomp}
\usepackage{xcolor}
\usepackage{gensymb}
\usepackage[colorlinks=true]{hyperref}

\newcommand{\modif}[1]{\textcolor{black}{#1}}

\begin{document}
\setcounter{secnumdepth}{4}
\title{Multiple metamagnetic transitions in helical antiferromagnet CeVGe$_3$}
\author{Hanshang Jin}
\affiliation{Department of Physics and Astronomy, University of California, Davis, California 95616, USA}
\author{Eun Sang Choi}
\affiliation{National High Magnetic Field Laboratory, Tallahassee, Florida 32310, USA}
\author{Hung-Cheng Wu}
\affiliation{Institute of Multidisciplinary Research for Advanced Materials, Tohoku University, Sendai 980-8577, Japan}
\affiliation{Department of Physics, National Sun Yat-sen University, Kaohsiung, 80424, Taiwan}
\author{N. J. Curro}
\affiliation{Department of Physics and Astronomy, University of California, Davis, California 95616, USA}
\author{K. Nawa}
\affiliation{Institute of Multidisciplinary Research for Advanced Materials, Tohoku University, Sendai 980-8577, Japan}
\author{T. J. Sato}
\affiliation{Institute of Multidisciplinary Research for Advanced Materials, Tohoku University, Sendai 980-8577, Japan}
\author{R. Kiyanagi}
\affiliation{J-PARC Center, Japan Atomic Energy Agency, Tokai, Ibaraki 319-1195, Japan}
\author{T. Ohhara}
\affiliation{J-PARC Center, Japan Atomic Energy Agency, Tokai, Ibaraki 319-1195, Japan}
\author{Peter Klavins}
\affiliation{Department of Physics and Astronomy, University of California, Davis, California 95616, USA}
\author{Valentin Taufour}
\email{vtaufour@ucdavis.edu}
\affiliation{Department of Physics and Astronomy, University of California, Davis, California 95616, USA}

\begin{abstract}
We report on neutron diffraction, magnetoresistance, magnetization and magnetic torque measurements under high magnetic field in the helical antiferromagnet CeVGe$_3$. This compound exhibits Kondo lattice coherence and helical antiferromagnetic (AFM) ordering at ambient pressure, similar to the well-studied CeRhIn$_5$. Our measurements reveal that CeVGe$_3$ undergoes a magnetic transition from an incommensurate (ICM) AFM state to an up-up-down-down commensurate (CM) AFM structure, followed by a transition to a novel phase at higher fields. \modif{A quantum phase transition occurs around 21.3\,T.} This rich magnetic field phase diagram closely resembles that of CeRhIn$_5$. Furthermore, angle-dependent magnetoresistance measurements reveal that all transitions in CeVGe$_3$ occur from the field component along the $ab$ plane. These findings highlight the intricate interplay among exchange interactions, crystal field effects, ground state properties, and crystalline symmetries.

\end{abstract}

\maketitle

\section{Introduction}
Strongly correlated electron systems represent a cornerstone of modern condensed matter physics, offering unique insights into the interplay between electron localization and itinerancy. These systems, which include high-temperature superconductors, iron-based superconductors, and heavy-fermion compounds, often exhibit rich emergent phenomena near \modif{quantum critical point (QCP)}~\cite{brando2016Metallic,stewart2001NonFermiliquid,lohneysen2007Fermiliquid,stewart2017Unconventional,aoki2013Heavy}. Heavy fermion materials, characterized by their enhanced effective electron masses, serve as prime examples of these complex interactions.

The tetragonal heavy fermion compound CeRhIn$_5$ has been extensively studied due to its elaborate phase diagrams under external parameters like pressure, doping, and magnetic fields~\cite{takeuchi2001Magnetic,muramatsu2001Superconductivity,knebel2011Antiferromagnetism,willers2015Correlation,jiao2015Fermi,ronning2017Electronic,rosa2019Enhanced,mishra2021Origin,mishra2021Specific}. At ambient pressure, this compound exhibits rich magnetic behaviors, transitioning from an incommensurate (ICM) antiferromagnetic (AFM) helical magnetic structure at low fields to a commensurate (CM) AFM structure at intermediate fields along the $ab$ plane, and entering into an electronic nematic phase~\cite{ronning2017Electronic,helm2020Nonmonotonic,kurihara2020Highfield}, or possibly another ICM AFM phase~\cite{mishra2021Origin,mishra2021Specific} at higher fields applied along the $c$ axis. More remarkably, CeRhIn$_5$ enters into a superconducting state under pressure around 2\,GPa, with a small domain of coexistence of AFM ordering and superconductivity~\cite{muramatsu2001Superconductivity,knebel2006Coexistence,knebel2011Antiferromagnetism}.

CeVGe$_3$ is a similar heavy fermion compound that also exhibits Kondo lattice coherence and helical antiferromagnetic order at ambient pressure below $T_\textrm{N} = 5.8$\,K, but it crystallizes in the hexagonal $P6_3/mmc$ space group. In this structure, the Ce occupies a site with $D_{3h}$~($\overline{6}m2$) point group symmetry and the ground state is a pure $\ket{\pm 1/2}$ doublet~\cite{inamdar2014Anisotropic,jin2022Suppression,chaffey2023Magnetic}. Both compounds have similar Kondo coherence temperatures~\cite{lin2015Evolution, chaffey2023Magnetic} and exhibit a comparable relocalization effect~\cite{chaffey2023Magnetic,shirer2012Long}. However, these compounds differ in their crystalline symmetries and Ce ground states, with CeRhIn$_5$ having a $\Gamma_7$ ground state, primarily of $\ket{\pm 5/2}$ state~\cite{willers2015Correlation}. Importantly, both materials exhibit a metamagnetic transition in magnetic fields applied perpendicular to the helical screw axis.

A previous NMR study suggested that the magnetic structure of the CeVGe$_3$ above the metamagnetic transition is an up-up-down-down CM AFM structure, similar to CeRhIn$_5$~\cite{chaffey2023Magnetic,fobes2018Tunable,mishra2021Origin}. In this study, we confirm this magnetic structure using high-field single-crystal neutron diffraction. We further explore the phase diagram of CeVGe$_3$ under high magnetic fields and discover a new first-order transition around 12\,T, similar to that of CeRhIn$_5$. A previous theoretical study proposed the mechanism for the helical magnetic order and the subsequent CM-AFM transitions in the presence of an intermediate in-plane field, and suggests that this higher field transition may involve a sinusoidal structure~\cite{nagamiya1962Magnetization}. The nature of this transition requires further investigation. \modif{A quantum phase transition (QPT) occurs at a magnetic field of approximately 21.3\,T.}

\section{Experimental Details}
\subsection{High Field Measurements}
The resistivity, magnetic torque, and magnetization at high fields above 30 T were measured at the National High Magnetic Field Laboratory in Tallahassee, FL. A conventional four-probe technique was used for the resistivity measurement with electrical contacts made on the $ab$ plane with silver epoxy. The dimension of the sample is 1.2 x 0.6 x 0.1 mm$^3$. The magnetic torque measurement was performed on a 1.7 mg single crystal sample with a capacitive torque magnetometer. The magnetic field was applied at a tilted angle from the $a$ axis by 5 degrees to induce a measurable torque signal. The magnetization was measured with a vibrational sample magnetometer (VSM) on a 30\,mg single crystal. \modif{Further details regarding sample preparation and quality can be found in Appendix~\ref{sample_quality}.}

\subsection{Neutron Scattering}
 To investigate magnetic structure at low temperatures and high fields, we conducted single-crystal neutron diffraction experiments at the time-of-flight (TOF) single-crystal neutron diffractometer SENJU, equipped with a vertical superconductor-7 T magnet, at J-PARC in Tokai, Japan~\cite{ohhara2016SENJU}. We selected two different incident neutron beams to cover the high and low Q ranges by using the first and second frames, respectively. The orientation of the single crystal is identified using GPTAS (4G) triple axis spectrometer~\cite{nawa2024Present} and a lab-based 4-circle XRD. The sample was mounted on an Al plate with the aid of a microscope so that $a$ axis is set almost vertical to the horizontal plane. During the experiment, it was found that the final condition had an insignificant misalignment of approximately 2.7 degrees, with the field almost parallel to the $a$ axis. Due to the $\pm 10\degree$ open angle restriction of the superconductor magnet, only horizontal detectors were employed in the experiment. The raw data was processed with the STARGazer program~\cite{ohhara2009Development} to obtain the squared structure factor ($|F|^2$) table for crystal structure and magnetic structure refinement.

\section{Results and Discussions}
\subsection{Magnetic structures at 0 T and 5 T}
\begin{figure}[!htb]
\centering
\includegraphics[width=0.65\linewidth]{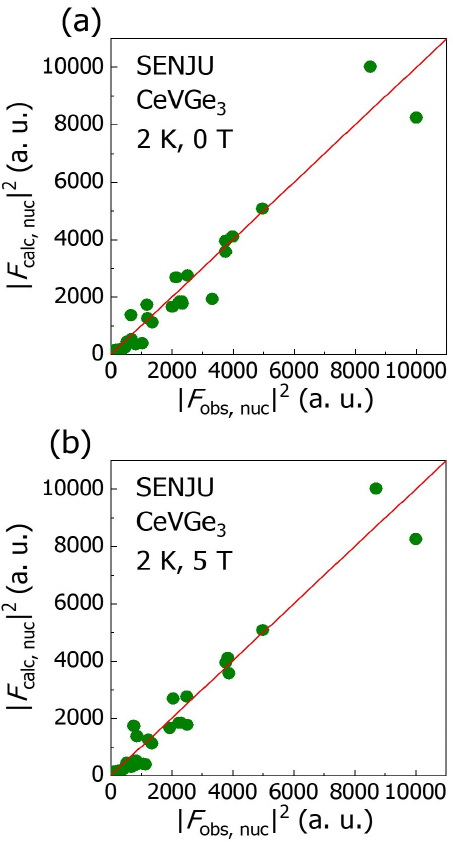}
\caption{Determination of the crystal structure at zero field and 5\,T  along the $a$ axis at 2\,K. The structure factors calculated from the refinement are compared to those observed experimentally at (a) 2\,K and 0\,T and (b) 2\,K and 5\,T. The results indicate that no structural change occurs between 0\,T and 5\,T.}
\label{Neutron_structure}
\end{figure}

The magnetic structure at zero field has been previously reported as an ICM AFM helical structure~\cite{chaffey2023Magnetic}. Consequently, the nuclear reflections and magnetic structure reflections do not overlap and can be analyzed separately. Initially, the crystal structure was identified at zero field and a temperature of 2\,K. A total of 28 nuclear reflections with $I$ $>$ 2$\sigma$ were analyzed, as summarized in Fig.~\ref{Neutron_structure}(a). The squared structure factor was numerically obtained from the integrated intensity using the following equation:
\begin{equation} \label{Nuc_intensity}
I_{cal}^{Nuc}=\frac{SV}{v_{0}^2}\frac{|F_N(hkl)|^2\lambda^4I(\lambda)}{sin^2\theta}
\end{equation}
where S, V, $v_0$, $F_N(hkl)$, $\lambda$, $\theta$, and $I(\lambda)$ represent the scale factor, volume of crystal, volume of the unit cell, nuclear structure factor, wavelength of incident neutrons, half of the scattering angle, and the wavelength dependent factor including the intensities of incident neutrons, respectively. Figure~\ref{Neutron_structure}(a) confirms that the crystal structure is $P6_3/mmc$, with no additional nuclear reflection observed, consistent with previous literature~\cite{chaffey2023Magnetic}.

\begin{figure}[!htb]
\centering
\includegraphics[width=0.65\linewidth]{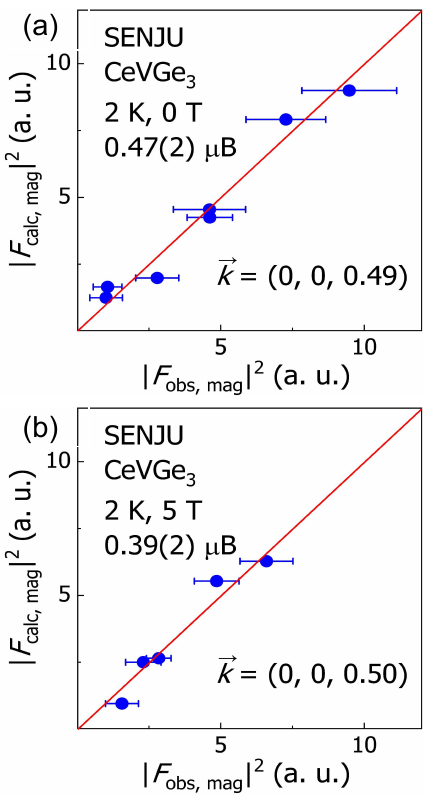}
\caption{Determination of the magnetic structure at both zero field and 5\,T along the $a$ axis at 2\,K. The results compare the magnetic structure factor calculated from the refinement with those observed experimentally at (a) 2\,K and 0\,T and (b) 2\,K and 5\,T. At zero field, the resolved magnetic structure is an incommensurate in-plane helical with a magnetic modulated propagation vector of $(0, 0, 0.49)$. At 5\,T, the magnetic structure changes to a commensurate up-up-down-down type with a magnetic modulated propagation vector of $(0, 0, 0.5)$. }
\label{Neutron_mag}
\end{figure}

Subsequently, the structural diffraction peaks were analyzed under an applied magnetic field of 5\,T along the $a$ axis. The absence of additional diffraction peaks indicates that the crystal structure remains $P6_3/mmc$ at 5\,T, as summarized in Fig.~\ref{Neutron_structure}(b). The scale factor at zero field and 5\,T was fixed to be the same. The zero field magnetic reflections, totaling 7 reflections with $I$ $>$ 2$\sigma$, were refined using the AFM helical model~\cite{chaffey2023Magnetic}, yielding a satisfactory correlation between the observed magnetic structure factor and the calculated magnetic structure factor, as shown in Fig.~\ref{Neutron_mag}(a). The square of the magnetic structure factor is numerically obtained from the integrated intensity using the following equation:
\begin{equation} \label{Mag_intensity}
I_{cal}^{Mag}=\frac{SV}{v_{0}^2}\frac{|F_M(hkl)|^2\lambda^4I(\lambda)}{sin^2\theta}
\end{equation}
where $F_M(hkl)$ represents the magnetic structure factor, which is proportional to square of the magnetic moment size ($m^2$).

The moment size of Ce$^{3+}$ at 2\,K was determined to be 0.47(2)\,$\mu_B$. At H = 5\,T, a clear shift in the magnetic reflections and the absence of two magnetic reflections indicate the emergence of a new magnetic phase at 5\,T. The magnetic modulation vector is indexed as $(0, 0, 0.5)$, corresponding to a commensurate phase. High field NMR results suggest the up-up-down-down commensurate model, characterized by the superposition of AFM and FM components~\cite{chaffey2023Magnetic}. Using the up-up-down-down model together with five measurable magnetic reflections, the refinement shows a good match with a reduced AFM moment size of 0.39(2)\,$\mu_B$, as summarized in Fig.~\ref{Neutron_mag}(b). The slight reduction in the magnetic moment can be attributed to the partial transformation of the magnetic moment from the AFM component perpendicular to the $a$ axis at zero field to the FM component along the $a$ axis at 5 T. It should be noted that the small FM component along the $a$ axis could not be detected in this experiment.

\begin{figure}
\centering
\includegraphics[width=\linewidth]{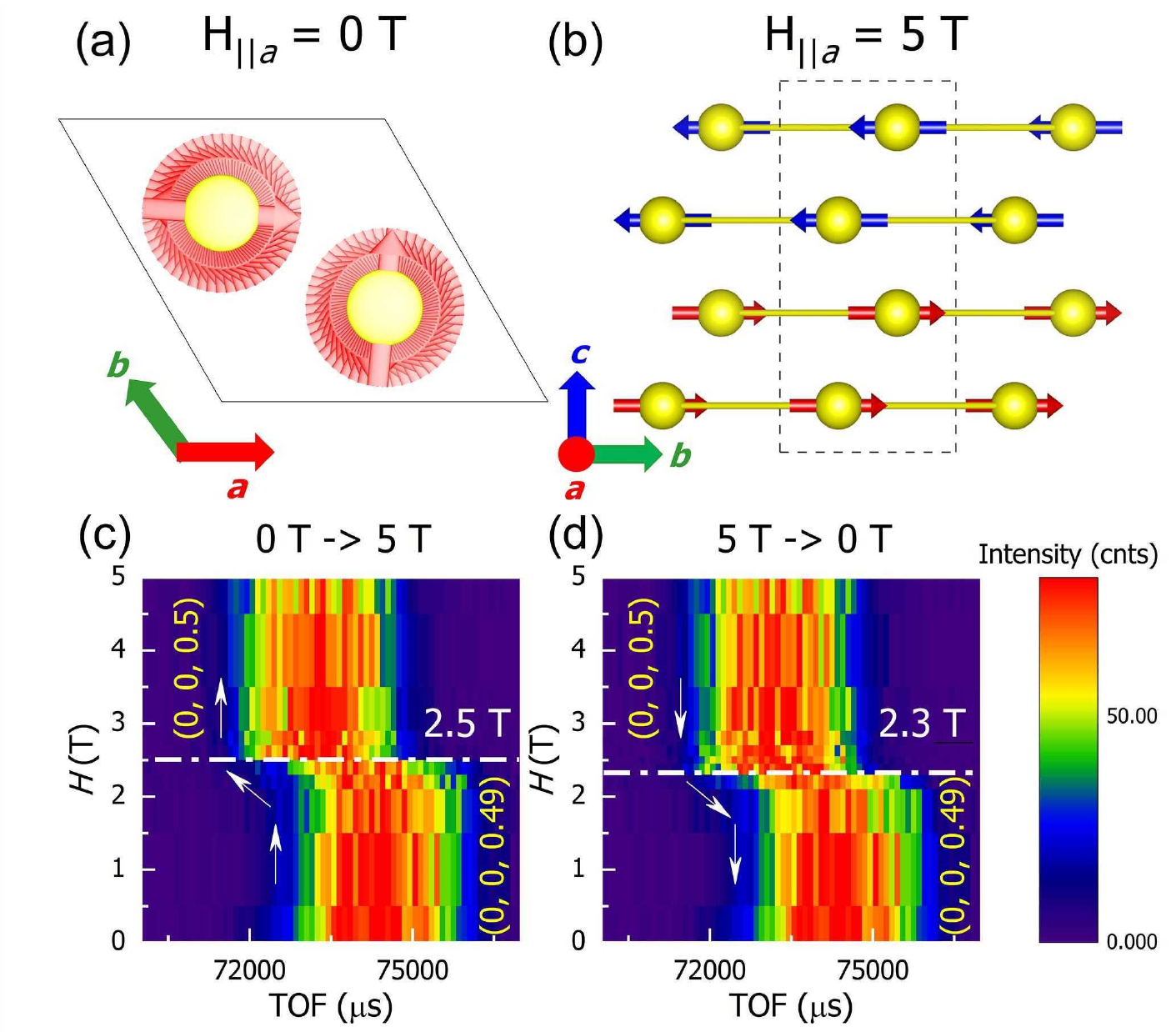}
\caption{The proposed magnetic structures at (a) zero field and (b) 5\,T. The field-dependent selected Q region at 2\,K demonstrates the transformation from ICM $(0, 0, 0.49)$ to CM $(0, 0, 0.5)$. (c) Field-increasing run from 0\,T to 5\,T, showing a transition field of 2.5\,T. (d) Field-decreasing run from 5\,T to 0\,T, showing a transition field of 2.3\,T.}
\label{mag_structure}
\end{figure}

A comparison of the magnetic structure at zero field and 5\,T is presented in Fig.~\ref{mag_structure}(a) and (b). Furthermore, the detailed field dependence of the selected Q region in the second frame is displayed in Fig.~\ref{mag_structure}(c) and (d). The TOF in 2D color map is proportional to the $d$-spacing and corresponds to a 1D line-scan along the 00$L$ direction. At zero field, the reflection at the higher TOF position is indexed as (0, 0, 0.49). Under a field of 5 T, the $d$-spacing of the reflection at the lower TOF position is twice that of the (0, 0, 1) reflection, indicating the development of (0, 0, 0.5) reflection. This clearly demonstrates that the (0, 0, 0.49) magnetic reflection remains unchanged below the transition field ($H_{1}$ = 2.5\,T) and suddenly shifts to the (0, 0, 0.50) position above $H_{1}$. This result reflects an incommensurate to commensurate transition in CeVGe$_3$. Additionally, a difference in $H_{1}$ is observed during the increasing and decreasing field runs, suggesting the presence of a hysteresis loop. \modif{We neglect the demagnetization effect on the average transition field because the sample is plate-like and the magnetic field is applied within the plane of the plate.}

\subsection{Magnetic Field Phase Diagram}

\begin{figure}[!htb]
\centering
\includegraphics[width=\linewidth]{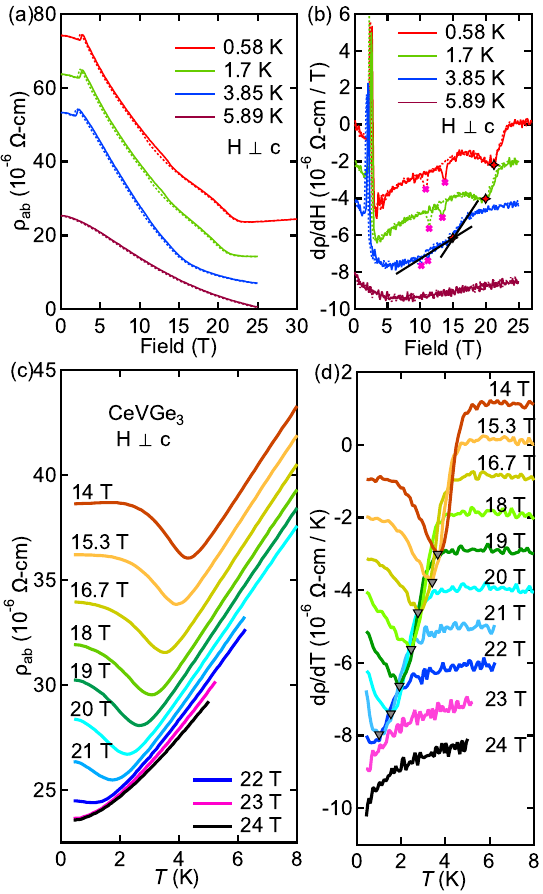}
\caption{(a) Selected field-dependent resistivity data with in-plane fields on CeVGe$_3$, and (b) their corresponding derivatives with consistent offsets of 10\,$\mu\Omega\,\mathrm{cm}$ and 2\,$\mu\Omega\,\mathrm{cm}\,\mathrm{T}^{-1}$ respectively to avoid overlapping. \modif{Solid lines represent the field-raising data, and the dashed lines represent the field-lowering data. They reveal a distinct hysteresis feature around 12\,T, and the hysteresis anomalies are marked as purple crosses. The transitions at the higher field are marked as red stars, and the quantum phase transition occurs around 21\,T.} (c) The temperature-dependent resistivity under high fields along the $ab$ plane and (d) their corresponding derivatives. The derivatives are vertically offset by 1\,$\mu\Omega\,\mathrm{cm}\,\mathrm{T}^{-1}$ to avoid overlapping.}
\label{Fig:HFdata}
\end{figure}

\begin{figure}[!htb]
\centering
\includegraphics[width=\linewidth]{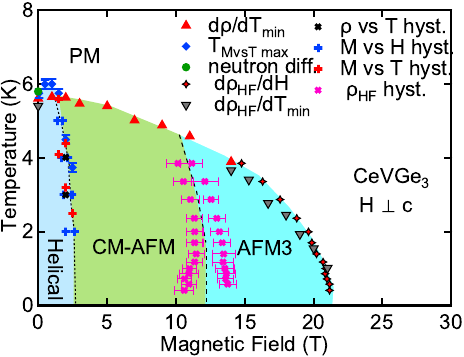}
\caption{H-T phase diagram of CeVGe$_3$ with in-plane applied fields. The red stars and purple crosses are from our preliminary high-field $\rho$ vs H data, and the grey triangles are the minimum of the derivatives of $\rho$ vs T data. \modif{The error bars for CM-AFM to AFM3 transitions are taken as half the peak width.} Other criteria are from Ref.~\cite{chaffey2023Magnetic}.}
\label{Fig:HFphase}
\end{figure}

The field and the temperature-dependent in-plane resistivity of a CeVGe$_3$ single crystal are shown in Fig.~\ref{Fig:HFdata}.
We conducted measurements of the field-dependent in-plane resistivity up to a maximum of 40\,T perpendicular to the $c$ axis and down to a minimum of 0.41\,K in temperature. Selected field-dependent resistivity data and corresponding derivatives are shown in Fig.~\ref{Fig:HFdata} (a) and (b). The anomalies around 2.5\,T correspond to the previously known metamagnetic transition for fields perpendicular to the $c$ axis~\cite{inamdar2014Anisotropic,chaffey2023Magnetic}. The field-raising and lowering magnetoresistance reveal a distinct hysteresis loop around 12\,T at low temperatures, indicating a potential new metamagnetic transition in that region. The derivative peaks within this region are marked with purple crosses and depicted in the magnetic field phase diagram in Fig.~\ref{Fig:HFphase}. We label these two regions by CM-AFM and AFM3.

The previous neutron scattering study concluded that the ground state of CeVGe$_3$ exhibits a single-$k$ ICM helical structure with $\vec{k} = (0, 0, 0.49)$~\cite{chaffey2023Magnetic}. Our neutron study confirms that the magnetic structure becomes the CM-AFM state with $\vec{k} = (0, 0, 0.50)$ at 5\,T. This ICM-CM transition around 2.5\,T closely resembles the transition observed in CeRhIn$_5$  at approximately 2\,T along the $a$ axis. In CeRhIn$_5$, the ICM state (AFM1) is characterized by a propagation vector of $\vec{k} = (0.5, 0.5, 0.298)$, whereas the CM state (AFM3) has a propagation vector of $\vec{k} = (0.5, 0.5, 0.25)$~\cite{raymond2007Magnetic}.

\begin{figure}[!htb]
\centering
\includegraphics[width=\linewidth]{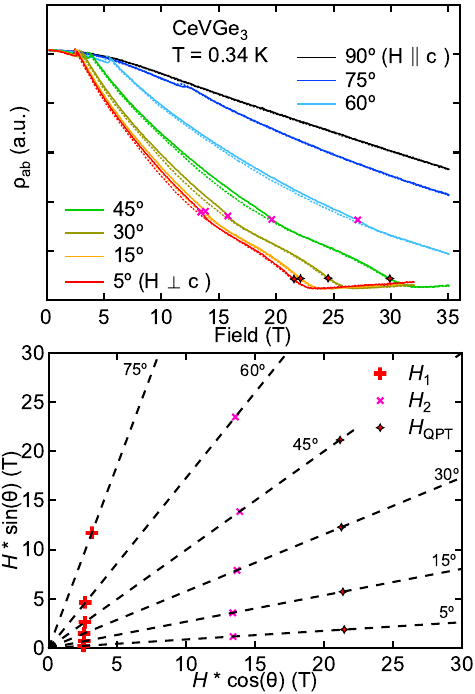}
\caption{(a) Field dependent in-plane resistivity at different fixed field angles in CeVGe$_3$ at T = 0.34\,K, up to 35\,T. \modif{Solid lines represent the field-raising data, and the dashed lines represent the field-lowering data. }(b) The \modif{transition} fields are determined based on the derivatives \modif{in field-raising data} and are plotted as their in-plane vs out-of-plane components. The \modif{transition} fields at different angles have consistent in-plane components, suggesting that the transitions occur exclusively within the $ab$ plane.}
\label{CeVGe3_angle}
\end{figure}

The angle-dependent in-plane magnetoresistance data are shown in Fig.~\ref{CeVGe3_angle} (a), and the transition fields, determined from the corresponding derivatives, are plotted as their in-plane versus out-of-plane components in Fig.~\ref{CeVGe3_angle} (b). The transition fields at different angles have consistent in-plane components, indicating that all transitions occur exclusively from the field component along the $ab$ plane, and are not affected by the field along the $c$ axis. This differs from the phase diagram of CeRhIn$_5$, where the low-field transition occurs with the applied field in-plane, and the high-field transition occurs with the applied field out-of-plane~\cite{mishra2021Specific}. \modif{The average in-plane component of the quantum phase transition field, $H_{QPT}$, is 21.3\,T.} 

We note that in the CM-AFM state, the moment lies in the $ab$ plane with the up-up-down-down configuration~\cite{fobes2018Tunable,mishra2021Origin} in both compounds, despite their different structural symmetries, ground states and orbital anisotropy. The different ground states in CeRhIn$_5$ (predominant $\ket{\pm 5/2}$~\cite{willers2015Correlation}) and CeVGe$_3$ (pure $\ket{\pm 1/2}$~\cite{jin2022Suppression}) primarily reflect variations in their crystal field environments. The closest ions to Ce atoms in CeVGe$_3$ are the Ge atoms within the same plane, leading to the ground state $\ket{\pm 1/2}$. The closest ions to Ce atoms in CeRhIn$_5$ are the In atoms with the Wyckoff position of $4i$ forming a square above and below the Ce, leading to the ground state that is mainly of $\ket{\pm 5/2}$.

\begin{figure}[!htb]
\centering
\includegraphics[width=\linewidth]{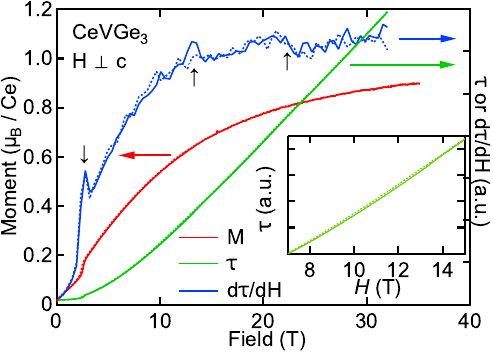}
\caption{Field-dependent magnetization \modif{at T = 1.5\,K} (red curve, left axis), magnetic torque \modif{at T = 0.34\,K}(green curve), and  \modif{its} derivative of the torque (blue curve). The solid(dash) curves are field-raising(lowering) data. The small arrows indicate the three transitions observed in $d\tau$/$d$\modif{$H$}. \modif{The inset highlights the hysteresis loop observed in magnetic torque near $H_{2}$.}}
\label{CeVGe3_MvsH}
\end{figure}

The field-dependent magnetization using the high-field VSM technique and the magnetic torque measurements are shown in Fig.~\ref{CeVGe3_MvsH}. \modif{The inset highlights the hysteresis feature in the magnetic torque.} Faint signals (indicated by small upward arrows) in the derivative of the torque might represent transitions from CM-AFM to AFM3 and across the \modif{QPT}. \modif{Although very small, the hysteresis disappears below $\sim7.9$\,T and above $\sim13.6$\,T which is consistent with the interpretation that the small hysteresis in torque measurements corresponds to the transition observed in resistivity $\rho(H)$.}
These transitions are not associated with the CEF level crossing, as the ground state is a pure $\ket{\pm 1/2}$ state, and a field along the $a$ axis only further favors the $\ket{\pm 1/2}$ state.
In CeRhIn$_5$, the anisotropic resistivity behavior is observed with a tilted field around 30\,T~\cite{ronning2017Electronic}. It is unclear whether this anisotropic resistivity behavior in CeRhIn$_5$ is due to the SDW. The anisotropic resistivity behavior previously believed to indicate a nematic state in Sr$_3$Ru$_2$O$_7$ with a similar tilted field around 8\,T has been shown to be an SDW state~\cite{lester2015Fieldtunable}.
The nature of the AFM3 state in CeVGe$_3$ requires further investigation, and high-field elastic neutron diffraction measurements will be helpful. NMR measurements in this phase also have the potential to identify this magnetic structure.

\section{Conclusion}
Our neutron scattering experiment confirms that the magnetic structure of CeVGe$_3$ above 2.5\,T is the up-up-down-down commensurate structure with a propagation vector of $(0, 0, 0.5)$.
The high field magnetoresistance measurements reveal a diversity of magnetic phases in CeVGe$_3$, transitioning from an ICM helical magnetic structure at low magnetic fields to a CM structure near 2.5\,T, entering into a new phase (AFM3) around 12\,T with a first-order transition, and \modif{a quantum phase transition occurs around 21.3\,T}. This rich magnetic field phase diagram closely resembles that of CeRhIn$_5$. Furthermore, angle-dependent magnetoresistance measurements reveal that all transitions in CeVGe$_3$ occur for the field applied within the $ab$ plane. These findings highlight the intricate interplay among exchange interactions, crystal field effects, ground state properties, and crystalline symmetries.

We hope our study will stimulate interest in theoretical and experimental works on the contrasting behaviors of these materials under various external conditions.

\section{Acknowledgement}
The SENJU experiment at the Materials and Life Science Experimental Facility of the J-PARC was performed under a user program (Proposals No. 2023B0319).
A portion of this work was performed at the National High Magnetic Field Laboratory, which is supported by the National Science Foundation Cooperative Agreement No. DMR-2128556 and the State of Florida.
Part of this work is financially supported by the Physics Department, University of California, Davis, U.S.A, and the NSF under Grant No. DMR-2210613.


\appendix
\section{Sample preparation and quality} 
\label{sample_quality}
\modif{Single crystals of CeVGe$_3$ were synthesized via the self-flux method, and the detailed procedure and temperature profile can be found in Ref.~\cite{chaffey2023Magnetic}. The starting materials [Ce pieces (Ames Lab), V pieces (etched with nitric acid), Ge lumps (6N)] were initially arc-melted to ensure a homogeneous mixture. The initial composition of elements is $\mathrm{Ce}:\mathrm{V}:\mathrm{Ge} = 4:1:19$. The arc-melted mixture was placed in a $2$\,mL Canfield Crucible Set~\cite{canfield1992Growth}, and sealed in a fused silica ampoule in a partial pressure of argon. The sealed ampoule was placed in a furnace where it was held at \SI{1200}{\celsius} for 10 hours, and slowly cooled to \SI{860}{\celsius} over 210 hours. At \SI{860}{\celsius}, the ampoule was removed from the furnace and quickly centrifuged to separate the single crystals from the molten flux.}

\begin{figure}[!htb]
\centering
\includegraphics[width=\linewidth]{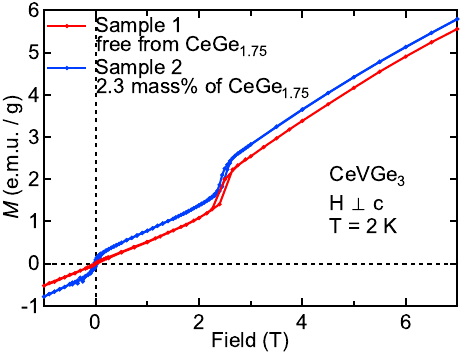}
\caption{\modif{Field-dependent magnetization of two different samples of CeVGe$_3$ at $T = 2$\,K. Sample \#1 shows no ferromagnetic jump in magnetization at low fields, indicating that the sample is free from the CeGe$_{1.75}$ impurity phase.}
\label{CeVGe3_smallsample}}
\end{figure}

\modif{The resistivity and magnetic torque measurements require relatively small samples. High-quality single crystals of CeVGe$_3$ were selected for these measurements based on the impurity content in the samples. Previous studies have reported that CeVGe$_3$ always contains some impurity phases, in particular trace amounts of CeGe$_{1.75}$~\cite{inamdar2014Anisotropic,chaffey2023Magnetic}, which tends to grow together within the CeVGe$_3$ crystals. CeGe$_{1.75}$ orders ferromagnetically below $7$\,K~\cite{budko2014Physical}. If the CeVGe$_3$ samples are small enough, it is possible to obtain pieces free from the CeGe$_{1.75}$ impurity phase. The sample quality can be assessed through the ferromagnetic signal attributed to the impurity in field-dependent magnetization measurements, as shown in Fig.~\ref{CeVGe3_smallsample}. A clean sample of CeVGe$_3$, like sample \#1, shows no ferromagnetic jump in magnetization at the low field. Based on the magnetization difference at 1\,T, sample \#2 contains an estimated 2.3\,mass\% of CeGe$_{1.75}$ impurity phase (we follow our estimation method described in~\cite{jin2021Stabilization}). However, this small amount of impurity does not affect the phase diagram, as the transitions from the helical to the CM-AFM state and from CM-AFM to AFM3 states can still be observed in CeVGe$_3$ with minor impurities during field-dependent resistivity measurements. The samples used in our high-field measurements are similar in mass and shape, and originated from the same batch as sample \#1.}

\begin{figure}[!htb]
\centering
\includegraphics[width=\linewidth]{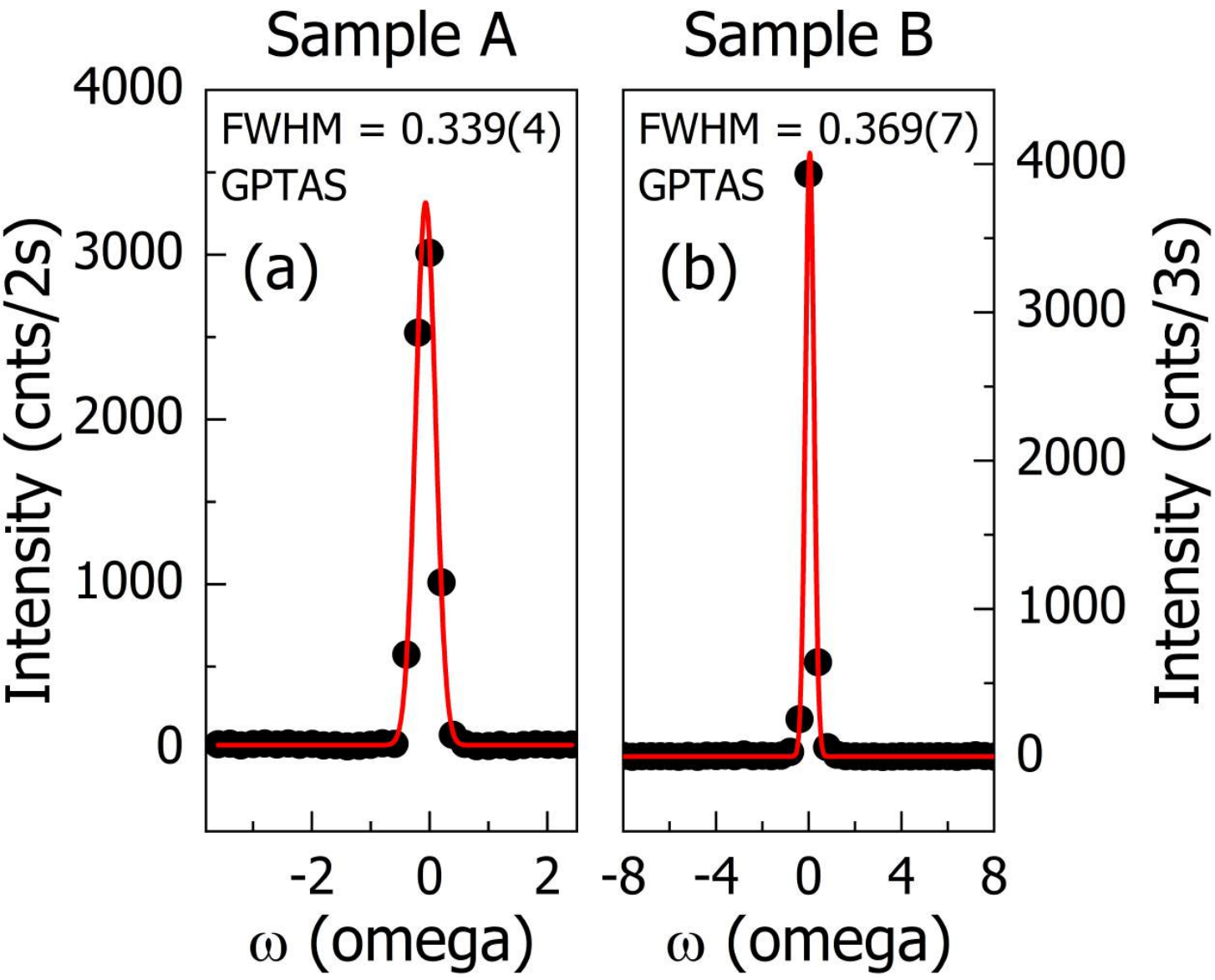}
\caption{\modif{Omega-scan at the 200 reflection shows a small mosaic spread for two large single crystals and confirms that both crystals have a single domain.}\label{CeVGe3_largesample}}
\end{figure}

\modif{The neutron scattering experiment and VSM measurements require relatively large samples to have better resolutions. Two large samples were characterized for the neutron scattering experiment. As shown in Fig.~\ref{CeVGe3_largesample}, the Omega-scan at the 200 reflection shows low mosaic spread for the two large single crystals and confirms that both crystals have a single crystallographic domain. Sample A was used for the neutron diffraction experiment, and sample B was used for the high-field VSM experiment.}

%

\end{document}